\documentclass{PoS}
\usepackage{subcaption}
\usepackage{wrapfig}
\usepackage{amsmath}

\title{Search for Primordial Black Hole Evaporation with VERITAS}

\ShortTitle{Search for PBH Evaporation with VERITAS}

\author{\speaker{Simon Archambault}\thanks{Now at Chiba University.}, for the VERITAS Collaboration\thanks{veritas.sao.arizona.edu}\\
        McGill University\\
        E-mail: \email{archs@chiba-u.jp}}

\abstract{Primordial black holes are black holes that may have formed from density fluctuations in the early universe. It has been theorized that black holes slowly evaporate. If primordial black holes of initial mass of $10^{14}$g were formed, their evaporation would end in this epoch, in a bright burst of very-high-energy gamma rays. A Cherenkov telescope experiment like VERITAS can look for these primordial black hole bursts in its archival data, constraining the rate-density of their final evaporation. New analysis techniques and search methodologies were used and will be presented here, leading to new constraints on the rate-density evaporation of primordial black holes, using 750 hours of archival VERITAS data.}

\FullConference{35th International Cosmic Ray Conference - ICRC217-\\
		10-20 July, 2017\\
		Bexco, Busan, Korea}

\begin{document}

\section{Primordial Black Holes}
\subsection{Description}
The idea behind black holes has been explored as early as in 1795, using Newtonian gravity. When the general theory of relativity came along, the theory was able to derive their properties using solutions to Einstein equations \cite{Schwarzschild1916}. Two types of black holes have been discovered, stellar-mass and supermassive, while two others have only been postulated: intermediate-mass and primordial black holes (PBHs).\\
PBHs were hypothesized by Stephen Hawking in 1971 \cite{Hawking1971}. In the early universe, density fluctuations are expected. Some of them might have been large enough to cause the matter in these locations to collapse due to the resulting gravitational pull, leading to the formation of primordial black holes. 

\subsection{Searches}
Hawking theorized that black holes evaporate \cite{Hawking1974}, by describing their thermodynamics, i.e. their entropy and temperature. In this treatment, black holes are expected to radiate particles (such as photons and pions) in a manner analogous to blackbody radiation, called Hawking Radiation. The lifetime $\tau$ of a black hole is expressed as $\tau\approx4.55\times10^{-28}(M_{BH}/1g)^3$ s \cite{MacGibbon1991}. This means that the evaporation time of the black hole is faster (to the power of 3) the lighter it is, leading to a burst of particles in the last moments of its lifetime. PBHs with a mass at formation of $\sim5\times10^{14}$ g would have an evaporation time of about the current age of the universe, meaning that the VHE component of their bursts might be seen now by Cherenkov Telescope experiments such as VERITAS.\\
Detecting such bursts would not only prove the existence of PBHs and determine their rate-density of evaporation, but also help in determining their relic density. This detection can also be used to study the power spectrum of the primordial density fluctuations \cite{Kim1999}, or to study their effect on the cosmic microwave background \cite{Naselskii1978}, on the creation of entropy \cite{Zeldovich1976}, baryogenesis \cite{Barrow1980} or nucleosynthesis \cite{Vainer1978}.\\
Previous experiments have searched for PBH evaporation. VERITAS has done such a search in the past and obtained a 99$\%$ CL limit on the rate-density of PBH evaporation of $1.29\times10^5$ pc$^{-3}$ yr$^{-1}$, when looking for bursts of a duration of 1 second \cite{Tesic2012}. The latest H.E.S.S. analysis looked at different burst durations, and found its best limits with one of 30 s, with a limit of $1.4\times10^4$ pc$^{-3}$ yr$^{-1}$ at the $95\%$ CL \cite{Glicenstein2013}. The Milagro experiment also obtained limits, being most sensitive with a burst duration of 1 second, for a limit of $3.6\times10^4$ pc$^{-3}$ yr$^{-1}$ \cite{Abdo2015}. 

\section{VERITAS}
\begin{wrapfigure}{L}{0.4\textwidth}
\centering
\includegraphics[width=0.38\textwidth]{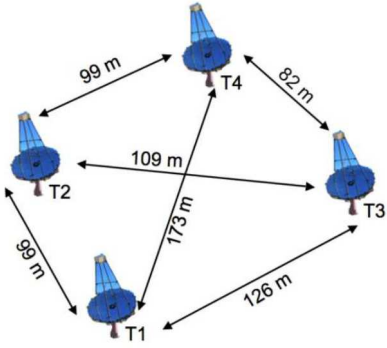}
\caption{Schematic of the VERITAS telescopes}
\label{VerSchem}
\end{wrapfigure}

The VERITAS array consists of four imaging atmospheric Cherenkov telescopes, located at the Fred Lawrence Whipple Observatory (FLWO) in southern Arizona (31 40N, 110 57W, 1.3km a.s.l.). A schematic of the array is shown in Figure \ref{VerSchem}. The telescopes follow a Davies-Cotton design, each with a 12-m tesselated mirror focussing Cherenkov light onto a camera comprising 499 PMTs. The field of view has a diameter of 3.5 degrees.

\section{Analysis}
The analysis is similar to what has been done in the past by VERITAS \cite{Tesic2012} and H.E.S.S. \cite{Abdo2015}, with some differences with regards to the understanding of the detector (like VERITAS' angular resolution, and the search for an optimal burst duration) as well as new analysis techniques recently made available with VERITAS (Boosted Decision Trees, or BDTs) \cite{Krause2017}.

\subsection{Boosted Decision Trees}
Data analysis done in this work used BDTs to do the gamma-hadron separation. In the PBH analysis shown in this work, moderate BDT cuts from \cite{Krause2017} were chosen. Moderate cuts are typically used when looking at sources following a power-law index of 2. The expected spectrum from PBH evaporation follows a power-law index of 1.5 (where hard BDT cuts would be used) up until $\sim$5-10 TeV, and then switches to an index of 3 (where soft BDT cuts would be used). This behavior is described in Equation \ref{PBHSpectrum}. Moderate cuts were determined to be a reasonable compromise.

\subsection{Data Selection}
The search for PBH bursts with VERITAS is done by looking at archival data. Since there is no preferred or expected time and direction for the final evaporation of a PBH, all data can be used, independent of the telescope pointings.
To optimize sensitivity, we use data taken in good weather and with all four telescopes in good working order. To achieve the lowest possible energy threshold, we require the telescope pointings to be above 50 degrees elevation. The data used here were taken before the upgrade of the VERITAS PMTs in the summer of 2012, but after the move of one of the telescopes to a location that improved the array sensitivity, in the summer of 2009. In future work, all of the data from all the VERITAS array configurations should be used, but for the scope of this work, this allows for a more direct comparison with previous VERITAS results \cite{Tesic2012}.

\subsection{Angular Resolution}
\begin{figure}
\centering
\includegraphics[width=0.6\textwidth]{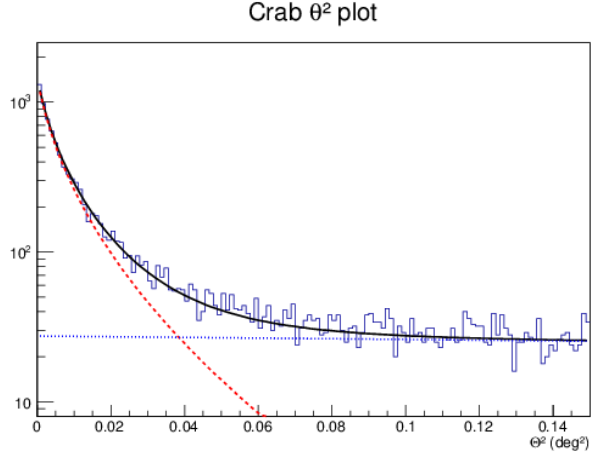}
\caption{$\Theta^2$-distribution of all the event from all the Crab runs, from all zenith angles and energies.}
\label{Theta2}
\end{figure}

The basic technique in the search for PBH evaporation is to look for bursts of gamma rays that come from the same direction, in a given duration. Therefore, it is necessary to understand the VERITAS point spread function (PSF).\\
VERITAS is often quoted, based on Monte Carlo simulations, to have a gamma-ray PSF of less than 0.1$^{\circ}$ at 1 TeV at the $68\%$ CL. Previous searches with VERITAS \cite{Tesic2012} and H.E.S.S. \cite{Glicenstein2013} used this PSF. However, the PSF will vary with energy and elevation, as the sizes of gamma-ray Cherenkov showers depend on those. This variation was determined by looking at the arrival direction of gamma rays from the Crab Nebula. The gamma-rays were categorized by energy and elevation, and their $\Theta^2$-distribution (the square of the angular distance between the target and the gamma-ray location) was drawn. The data were fit using a modified hyperbolic secant distribution, determined empirically \cite{Meagher2015}:

\begin{equation}
\label{hypSec}
S(\Theta^2, w)~=~\frac{1.71N}{2\pi w^2}sech(\sqrt{\Theta^2}/w)
\end{equation}

where w is the width of the distribution, representing 55.1$\%$ containment radius, and N is the number of signal events. Figure \ref{Theta2} shows a plot of all the data fit with this function (added to a constant background contribution). To measure the energy and elevation dependence of the PSF, separate fits were performed to subsets of the data.

\begin{figure}
\centering
\begin{minipage}{.5\textwidth}
\centering
\includegraphics[width=0.8\linewidth]{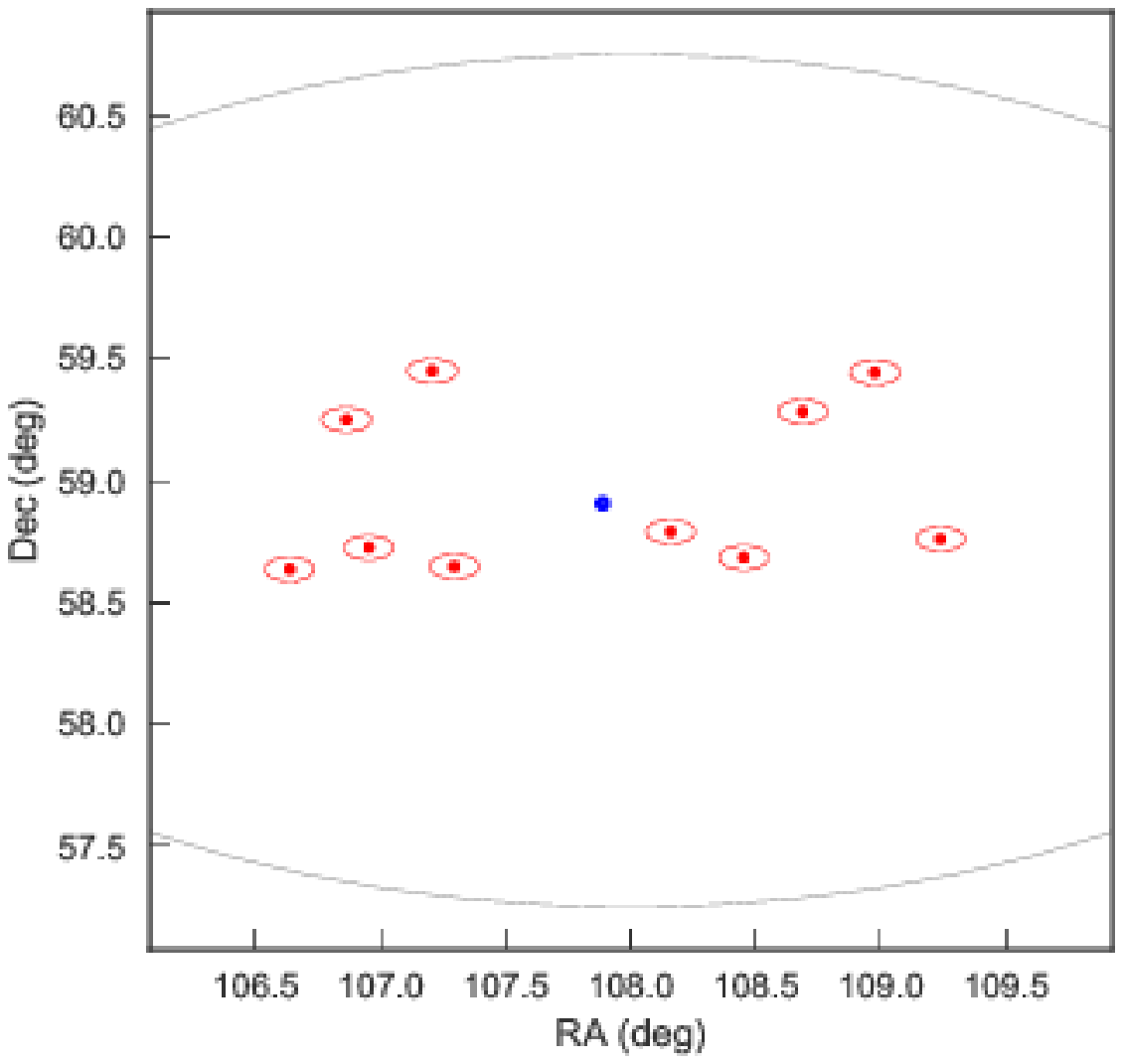}
\end{minipage}%
\begin{minipage}{.5\textwidth}
\centering
\includegraphics[width=0.8\linewidth]{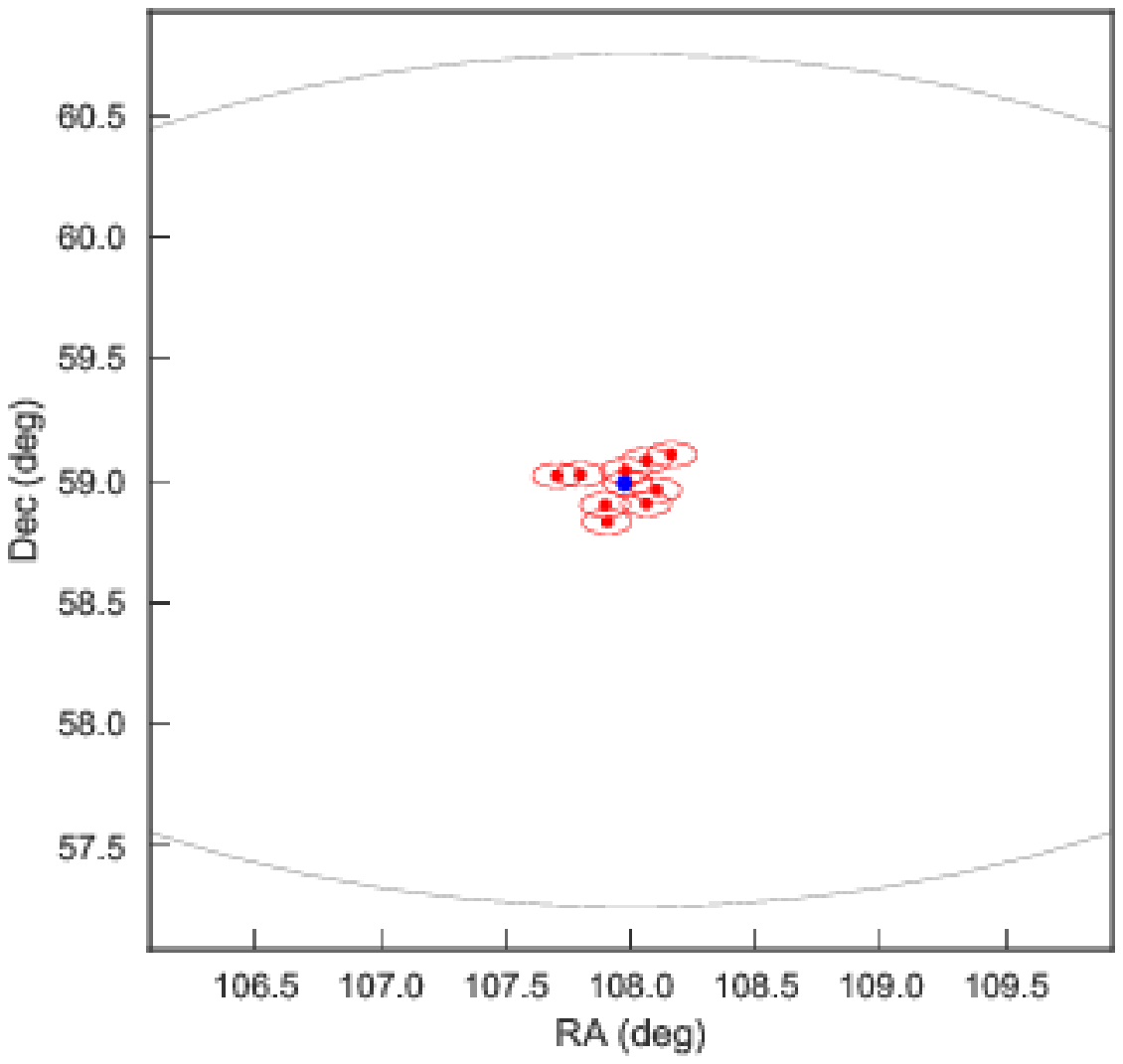}
\end{minipage}
\caption{\textit{Left:} Event centroid for a random group of events. \textit{Right:} A group of events whose positions were randomly generated from Equation \ref{hypSec}, to simulate a burst. The circles around the red points are the errors on the reconstructed event positions, while the blue point is the centroid of its respective burst. The random events do not fit with the centroid on the left plot, while they are highly-concentrated around the centroid on the right plot. The ellipses represent the field-of-view of the VERITAS telescopes, converted from camera to equatorial coordinates}
\label{centroid}
\end{figure}

As many more events are expected at lower energies or lower elevation, the selected binning has to be wider at high energies in order to get enough statistics. The energy binning chosen is the same one as what was used in the development of the VERITAS BDTs \cite{Krause2017}, i.e. 0.08 to 0.32 TeV, 0.32 to 0.5 TeV, 0.5 to 1 TeV, and 1 to 50 TeV, and 50 to 70 degrees in elevation, 70 to 80 degrees, and 80 to 90 degrees.\\
 The width $w_i$, which represents the angular resolutoin, is then used in a maximum-likelihood method to calculate the centroid of a given set of events. The idea is to use this to determine whether a set of events is likely to be coming from a point source. The centroid is determined by using this likelihood function:

\begin{equation}
L~=~\prod_i\frac{1.71}{\pi w_i^2}sech(\sqrt{(\Theta_i-\mu)^2}/w_i)
\end{equation}

where $w_i$ and $\Theta_i$ are the angular resolution and direction of event $i$ respectively, the width being given from the event's energy and elevation. Figure \ref{centroid} shows the centroid found by the minimization from a randomized set of events from a given run (to simulate background), as well as one from a set of events whose positions were randomly generated using Equation \ref{hypSec} (to simulate signal).  

\begin{wrapfigure}{R}{0.5\textwidth}
\centering
\includegraphics[width=0.48\textwidth]{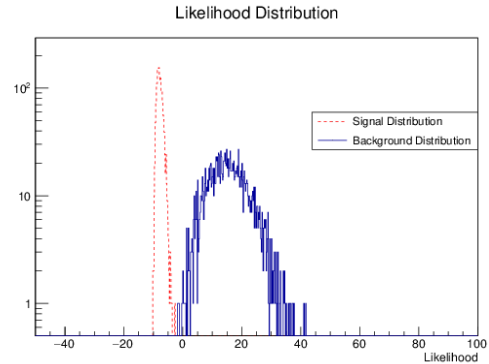}
\caption{Distribution of simulated signal bursts compared with random groups, of 5 events each, to determine cut values to use in the analysis.}
\label{likeDistr}
\end{wrapfigure}

Figure \ref{likeDistr} shows a likelihood distribution for groups of 5 events. The blue curve shows the likelihood distribution of random groups of 5 events, as background, while the red curve shows for 5-event simulated signal bursts. A cut is determined to be where $90\%$ of the signal bursts are included, in an effort to maximize the amount of signal bursts while minimizing the amount of background contamination. This exercise is done for bursts of 2, 3, 4, ..., 10 events, after which the likelihood of finding bursts with so many events becomes low, and the cuts tend to stay fairly constant between burst sizes.

\subsection{Burst-finding algorithm}
\begin{figure}
\centering
\includegraphics[width=0.6\textwidth]{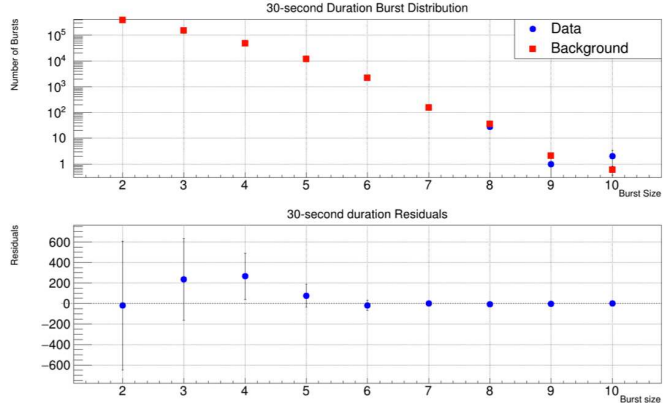}
\caption{Burst distribution for a burst duration of 30 seconds. \textit{Top: }The blue circles are from the data, and the red squares are the background estimation. \textit{Bottom: }Difference between data and background.}
\label{burstData}
\end{figure}

The procedure to find bursts in data is done on individual data runs. For each run, a list of events is compiled, flagged as either gamma-like or not using BDT cuts. For each events arriving at a time $t_i$, a list is compiled with any subsequent event arriving at a time $\Delta t$ later ($\Delta_t$ used in this analysis were 1, 2, 5, 10, 30 and 45 seconds). In these lists, only the events which are determined as likely to come from a point source are kept. This is done by finding the centroid of the events, and calculating their likelihood values. If the likelihood value is less than the cut determined from the method described in the previous section, all the events are kept as one burst. Otherwise, the event which is considered the `worst-fit' is removed, and the centroid and likelihood tests are done again using the remaining events. This is repeated until only one event is left, in which case the algorithm tries again with the discarded events, to catch other potential combinations that may have been missed. In the case that an event is found in more than one burst, it is classified as a member of the largest burst it is in, to avoid double-counting.\\
The methodology to describe the background is done the same way that was used in previous searches \cite{Tesic2012, Glicenstein2013}. The background is estimated by taking all the events of each run by using the same time stamps but assigning them to different events (to preserve trigger rates and dead times). This will break up any bursts (real signal bursts or accidental) in the data, and create new purely accidental ones. This last process is repeated 10 times to average out any statistical fluctuation, and gives an estimate of the number of expected, accidental bursts. This also presents the advantage of taking into account instrumental effects; for example, if a bright star was present during data-taking, its effects will carry over in the background estimation. Figure \ref{burstData} shows a burst distribution found using this algorithm for a time-window of $30$ seconds. The top plot compares the data points with the background estimation, while the lower plot shows the difference.

\section{Results}
Limits are derived by using a maximum likelihood method, looking for a 99$\%$ CL for the rate-density of PBH evaporation based on the number of signal and background events.  The likelihood equation is as follows:

\begin{equation}
-2\ln L~=~\sum_b n_b * \ln(n_{excess} + n_{bg}) - n_{excess}
\end{equation}

where $b$ is the size of the burst (or the number of events in a burst), $n_b$ is the number of bursts of size $b$ in data, $n_{bg}$ is the number of bursts of size $b$ in the background estimation, and $n_{excess}$ is the number of excess events. The number of excess events is the quantity to minimize, but is determined by the expected number of bursts that should be seen given the rate-density of PBH evaporation. It also depends on the emission model. The one used in this work is the same model used by Milagro in \cite{Ukwatta2016}:

\begin{equation}
\label{PBHSpectrum}
\begin{split}
\frac{dN_\gamma}{dE_\gamma}&\approx~9\times10^{35}\\
&\times\begin{cases}
\left(\frac{1\text{GeV}}{T_\tau}\right)^{3/2}\left(\frac{1\text{GeV}}{E_\gamma}\right)^{3/2}~\text{GeV}^{-1}&\text{for}~E_\gamma<kT_\tau\\
\left(\frac{1\text{GeV}}{E_\gamma}\right)^3~\text{GeV}^{-1}&\text{for}~E_\gamma\geq kT_\tau\\
\end{cases}
\end{split}
\end{equation}

where $T_\tau$ is the temperature of the black hole at the beginning of the final burst time interval, defined as $kT_\tau=7.8(\tau/1s)^{-1/3}$ TeV. The number of expected photons $N_\gamma$ seen from a PBH burst would be expressed as:

\begin{equation}
N_\gamma(r, \alpha, \delta, \Delta t)~=~\frac{1}{4\pi r^2}\int_0^\infty\frac{dN}{dE}(E_\gamma, \Delta t)A(E_\gamma, \theta_z, \theta_w, \mu, \alpha, \delta)dE_\gamma
\end{equation}

where $A(E_\gamma, \theta_z, \theta_w, \mu, \alpha, \delta)$ is the instrument response function (effective area and camera radial acceptance) of VERITAS, as a function of the gamma-ray energy $E_\gamma$, the observation zenith angle $\theta_z$, the source's offset from the center of the camera (or wobble) $\theta_w$, the optical efficiency $\mu$, and the event reconstruction position in camera coordinates $(\alpha, \delta)$.\\
The probability of seeing a burst of a certain number of photons $b$ from a PBH emitting $N_\gamma$ VHE photons is expressed through Poisson statistics: $P_{(b, N_\gamma)}=e^{-N_\gamma}N_\gamma^b/b!$. The effective volume $V_{eff}$ is then calculated this way:

\begin{equation}
V_{eff}(b, \Delta_t)~=~\int_{\Delta\Omega} d\Omega\int_0^\infty drr^2P(b, N_\gamma)
\end{equation}

Finally, this means that the expected number of bursts seen by VERITAS is expressed like this:

\begin{equation}
n_{exp}(b, \Delta t)~=~n_{excess}~=~\dot\rho_{PBH}\times T_{obs} \times V_{eff}(b,\Delta t)
\end{equation}

where $\dot\rho_{PBH}$ is the rate-density of PBH evaporation, and $T_{obs}$ is the duration of observations.

\begin{figure}
\centering
\begin{minipage}{.5\textwidth}
\centering
\includegraphics[width=0.8\linewidth]{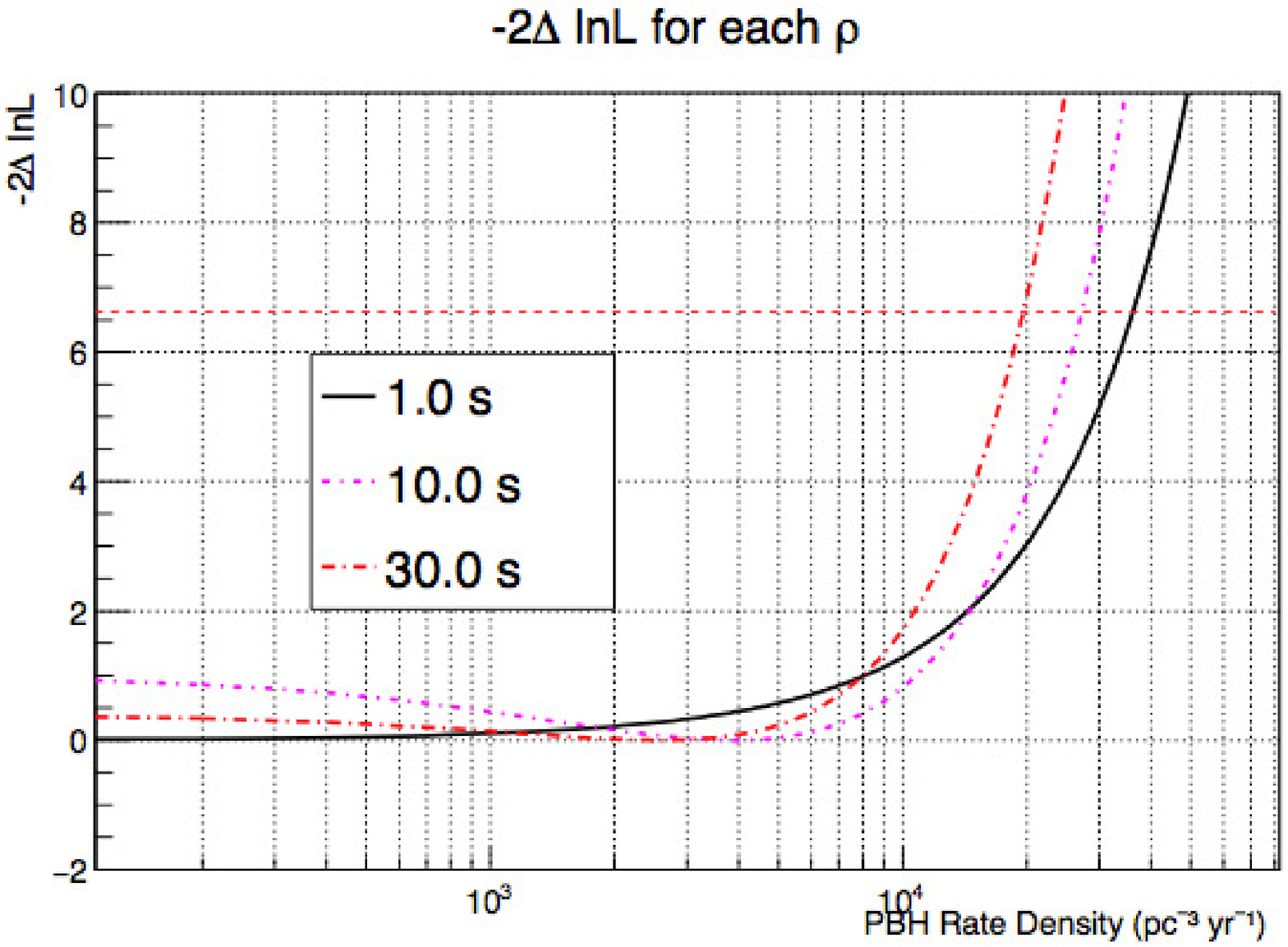}
\end{minipage}%
\begin{minipage}{.5\textwidth}
\centering
\includegraphics[width=0.8\linewidth]{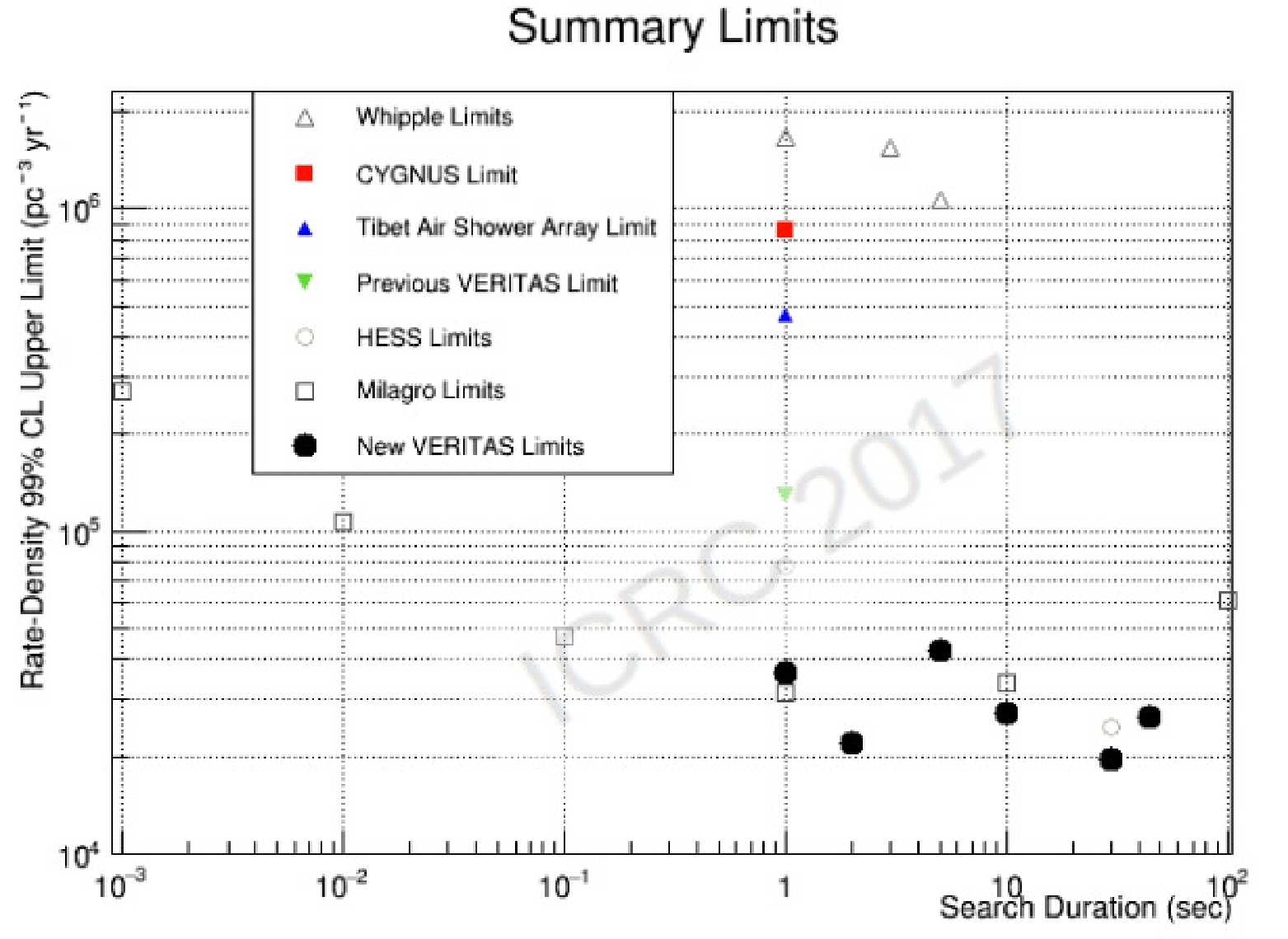}
\end{minipage}
\caption{\textit{Left:} Likelihood curves for three of the burst durations used in this analysis, at 1, 10 and 30 seconds. \textit{Right:} Summary of the limits found in this analysis, compared to results from previous experiments (numbers taken from \cite{Ukwatta2016}).}
\label{LikeRes}
\end{figure}

 The 99$\%$ CL limit is found when $-2\Delta\ln L \leq 6.63$, where $\Delta\ln L=\ln L(n_b|\dot\rho_{PBH})-\ln L_{min}$. The left plot of Figure \ref{LikeRes} shows the likelihood curve for different durations analyzed here, where the dashed line at 6.63 indicates the 99$\%$ CL. The right plot shows the limits for the different burst durations, compared to measurements from previous searches. The optimal duration is found using a polynomial fit of the limits, and is around a $30$s time-window, for 99$\%$ CL limit of $2.22\times10^4$ pc$^{-3}$ yr$^{-1}$, a factor of $1.5$ better than the previous best limits set by Milagro, with a time-window of 1 second \cite{Abdo2015}. These were obtained using 747 hours of data.

\section{Summary}
New limits were obtained on the rate-density of PBH evaporation with VERITAS, using boosted decision trees (BDTs) for gamma-hadron separation, as well as by taking into account the behavior of the VERITAS PSF as a function of energy and elevation. The best limits obtained were at a duration of $30$ s, for a rate-density of $2.22\times10^4$ pc$^{-3}$ yr$^{-1}$, using 747 hours of data.

\acknowledgments
This research is supported by grants from the U.S. Department of Energy Office of Science, the U.S. National Science Foundation and the Smithsonian Institution, and by NSERC in Canada. We acknowledge the excellent work of the technical support staff at the Fred Lawrence Whipple Observatory and at the collaborating institutions in the construction and operation of the instrument. The VERITAS Collaboration is grateful to Trevor Weekes for his seminal contributions and leadership in the field of VHE gamma-ray astrophysics, which made this study possible.

\end{document}